\documentclass[useAMS,usenatbib]{biom}

\def\bSig\mathbf{\Sigma}

\usepackage{amsmath}
\usepackage{graphicx}
\usepackage{multirow}
\usepackage{xcolor}
\usepackage{tikz}

\newcommand{\bl}[1]{\textcolor{black}{#1}}

\DeclareMathOperator{\expit}{expit}

\usetikzlibrary{decorations.pathreplacing}
\usetikzlibrary{arrows.meta}

\usetikzlibrary{plotmarks}

\title[Bayesian Causal Analysis of Recurrent Events]{A Bayesian Framework for Causal Analysis of Recurrent Events with Timing Misalignment}

\author{Arman Oganisian$^{1,*}$\email{arman\_oganisian@brown.edu}, 
Anthony Girard$^{2}$, Jon A. Steingrimsson$^{1}$,and Patience Moyo$^{3}$ \\
$^{1}$Department of Biostatistics, Brown University, Providence, Rhode Island, U.S.A. \\
$^{2}$Leonard Davis Institute for Health Economics, University of Pennsylvania, Philadelphia, Pennsylvania, U.S.A. \\
$^{3}$Department of Health Services, Policy, and Practice, Brown University, Providence, Rhode Island, U.S.A.}

\begin{document}

\date{{\it Received July 2023 }  {\it
Accepted November 2024} }

\pagerange{\pageref{firstpage}--\pageref{lastpage}} 
\volume{(In Press)}
\pubyear{2024}
\artmonth{Accepted November}

\doi{10.1093/biomtc/ujae145}

\label{firstpage}

%  put the summary for your paper here

\begin{abstract}
Observational studies of recurrent event rates are common in biomedical statistics. Broadly, the goal is to estimate differences in event rates under two treatments within a defined target population over a specified followup window. Estimation with observational data is challenging because while membership in the target population is defined in terms of eligibility criteria, treatment is rarely observed exactly at the time of eligibility. Ad-hoc solutions to this timing misalignment can induce bias by incorrectly attributing prior event counts and person-time to treatment. Even if eligibility and treatment are aligned, a terminal event process (e.g. death) often stops the recurrent event process of interest. In practice, both processes can be censored so that events are not observed over the entire followup window. Our approach addresses misalignment by casting it as a time-varying treatment problem: some patients are on treatment at eligibility while others are off treatment but may switch to treatment at a specified time - if they survive long enough. We define and identify an average causal effect estimand under right-censoring. Estimation is done using a g-computation procedure with a joint semiparametric Bayesian model for the death and recurrent event processes. We apply the method to contrast hospitalization rates among patients with different opioid treatments using Medicare insurance claims data.\end{abstract}


%  As usual, the \maketitle command creates the title and author/affiliations
%  display 

\maketitle

\section{Introduction} \label{sc:intro}

The increasing availability of large medical insurance claims data bases and electronic health records has popularized observational studies of recurrent event outcomes. The goal of these analyses is generally to contrast event rates within a target population over a given followup window under hypothetical treatment interventions (hereafter, treated and untreated). Causal recurrent event modeling using observational data poses two unique sets of challenges. 

The first set is related to causal modeling. While several recurrent events can be observed, the recurrent event process may be terminated by an event such as death. Additionally, a censoring process coarsens the data up to a lower bound on both survival time and the accumulated number of events. Moreover, event rates are composite outcomes with survival time in the denominator and event counts in the numerator. Since both are potential outcomes of treatment, they must be modeled jointly over time as a function of treatment.

A second set design difficulties involves choosing when to begin followup. Principles of target trial emulation \citep{Hernan2016} suggest that followup start when 1) eligibility criteria are met, 2) treatment strategies are assigned, and 3) the recurrent/terminal events begin to be counted. The eligibility criteria define membership into the target population, which is crucial for interpretation. The design issue arises when patients rarely receive treatment exactly at the time of eligibility. This timing misalignment leads to an array of ad-hoc strategies of determining treatment assignment that may induce bias. These include looking past eligibility to check whether a patient was ever treated and, if so, classifying them as being in the treatment group. A comparison of post-eligibility event rates between groups will in general be biased: in the constructed treated group, both the person-time and event counts accrued between eligibility and treatment initiation are misattributed to treatment. Alternatively, analyzing events starting from time zero being treatment initiation may induce survivor bias as it shifts the target population to the subpopulation who survived long enough to start treatment. If survival is differential across treatments, this could lead to bias in the treatment effect estimate. In our data analysis, for instance, we are concerned with estimating the effect of opioid medication on hospitalization risk among patients with chronic back pain using Medicare insurance claims data. Patients are included in the study based on several eligibility criteria - such as presence of chronic pain diagnosis codes. Follow-up begins when these diagnosis criteria are met but typically only very few subjects will have had opioid medication filled exactly at the time of diagnosis. Therapy may begin later in the followup period, provided the patient survives long enough. 

Our contributions advance existing work along several important fronts. Misalignment of study and treatment starts is well-studied with survival outcomes, where bias due to misattribution of the person-time between study and treatment start is termed immortal time bias. Various methods such as grace periods and landmarking have been proposed \citep{Duchesneau2022} to correct for this bias. While precise causal framing of immortal time bias, grace periods \citep{Wanis2022}, and solutions such as cloning \citep{Hernan2018} have been formulated, direct application to recurrent event data is not appropriate since, as we show, recurrent event outcomes require a separate set of identification assumptions, estimands, and models.
 
Early developments in recurrent event models include proportional mean models \citep{Ghosh2002}, frailty models \citep{Cowling2006}, and multistate model \citep{andersen2002}. Recently, interest has grown in developing a precise causal framing of recurrent events. For instance, one proposal \citep{Su2020a} used a set of estimators of causal effects on cumulative event rate over a specified time. However, unlike our approach, it does not consider terminal events such as death within followup. Inverse-probability weighted estimators \citep{Schaubel2010} with doubly-robust extensions \citep{Su2020b} have also been developed. However, these approaches are developed in the point-treatment setting where there is no timing misalignment. Other work \citep{Lyu2022} developed Bayesian estimation for a difference in average potential number of events within a principal stratum of survivors, but only in a point-treatment setting. Additionally, a drawback of principal stratification is that estimands are conditional on a non-identifiable stratum membership, leading to difficulty in interpretation. In contrast, our work targets a marginal estimand with a clear causal interpretation. Recent work \citep{Janvin2024} reviews causal estimands in recurrent events and provides a discrete-time causal framework but in a point-treatment setting.

We develop an all-in-one Bayesian approach that simultaneously 1) handles timing misalignment, 2) is valid under covariate-dependent right-censoring, 3) accounts for terminal processes while modeling recurrent event counts, 4) adjusts for measured confounding within a formal causal inference framework. Specifically, we discretize the followup and define a time-varying indicator for each patient that is zero before switching to treatment and one afterwards. The causal target estimand of interest is the difference in average potential incidence rates had subjects switched to treatment at interval $s$ versus $s'$. This reframes the issue as a time-varying treatment problem, which we solve with an appropriately constructed g-method \citep{robins1986}. Under assumptions on the joint treatment-censoring mechanism, we derive a g-formula that nonparametically identifies our estimand in terms of the joint distribution of survival and event counts. We model these using a sequence of two semiparametric Bayesian models: a discrete-time survival model and a model for the count of events in a time interval given survival up to that interval. Though standard when implementing g-methods, discrete-time models are difficult to fit with standard frequentist methods. For instance, \cite{Young2020} document having to simplify the models as ``the construction of confidence intervals failed to converge under the more flexible model'' in the required bootstrap procedures. The post-selection inference issues of iterating between many different models until convergence is attained may compromise the validity of frequentist inference. Discarding the pathological bootstrap iterations is also not proper. In contrast, our Bayesian approach uses autoregressive priors that regularize the models while full posterior inference avoids irregularities associated with the frequentist bootstrap. In the subsequent sections, we outline the notation and observed data structure, specify and identify the causal estimand, and present our Bayesian models. We conduct simulation experiments exploring frequentist properties and end with an analysis of opioid use and hospitalization using Medicare data.

\begin{figure*}[t]
          \centering
          \resizebox{.8\linewidth}{!}{\begin{tikzpicture}

%------------------- Axis  -----------------------%
\draw (0,-5) -- (10,-5);
\draw (5,-5.2) node[below, yshift=-2.5em]{time};
\draw (0,-5) node[below, yshift=-.1em, xshift=-.7em]{$\tau_0=0$};
\draw (10, -5) node[below, yshift=-.1em, xshift=1em]{$\tau_{10}=\tau$};
\draw [dashed] (10,-5) -- (10,.5);

%------------------- Partition -------------------%
\draw [dashed] (9,-5) -- (9,.5);
\draw [dashed] (8,-5) -- (8,.5);
\draw [dashed] (7,-5) -- (7,.5);
\draw [dashed] (6,-5) -- (6,.5);
\draw [dashed] (5,-5) -- (5,.5);
\draw [dashed] (4,-5) -- (4,.5);
\draw [dashed] (3,-5) -- (3,.5);
\draw [dashed] (2,-5) -- (2,.5);
\draw [dashed] (1,-5) -- (1,.5);

\draw (1,-5) node[below, yshift=-.1em]{$\tau_1$};
\draw (2,-5) node[below, yshift=-.1em]{$\tau_2$};
\draw (3,-5) node[below, yshift=-.1em]{$\tau_3$};
\draw (4,-5) node[below, yshift=-.1em]{$\tau_4$};
\draw (5,-5) node[below, yshift=-.1em]{$\tau_5$};
\draw (6,-5) node[below, yshift=-.1em]{$\tau_6$};
\draw (7,-5) node[below, yshift=-.1em]{$\tau_7$};
\draw (8,-5) node[below, yshift=-.1em]{$\tau_8$};
\draw (9,-5) node[below, yshift=-.1em]{$\tau_9$};

\draw (1,-5) node[below, yshift=-1em, xshift=-1em]{$I_1$};
\draw (2,-5) node[below, yshift=-1em, xshift=-1em]{$I_2$};
\draw (3,-5) node[below, yshift=-1em, xshift=-1em]{$I_3$};
\draw (4,-5) node[below, yshift=-1em, xshift=-1em]{$I_4$};
\draw (5,-5) node[below, yshift=-1em, xshift=-1em]{$I_5$};
\draw (6,-5) node[below, yshift=-1em, xshift=-1em]{$I_6$};
\draw (7,-5) node[below, yshift=-1em, xshift=-1em]{$I_7$};
\draw (8,-5) node[below, yshift=-1em, xshift=-1em]{$I_8$};
\draw (9,-5) node[below, yshift=-1em, xshift=-1em]{$I_9$};
\draw (10,-5) node[below, yshift=-1em, xshift=-1em]{$I_{10}$};

%------------------- Subject 1 -----------------------%
\draw (-1.5,.5) node[below, yshift=-.6em]{Subject 1};
\draw (0,0pt) -- (7.2,0pt);

\draw (0,3pt) -- (0,-3pt) node[below, xshift = 10pt]{};

\draw (1.5,3pt) -- (1.5,-3pt) node[below]{$V_1$};
\draw (2.4,3pt) -- (2.4,-3pt) node[below]{$V_2$};
\draw (3.75,3pt) -- (3.75,-3pt) node[below]{$V_3$};
\draw (4.5,3pt) -- (4.5,-3pt) node[below]{$V_4$};
\draw (5.5,3pt) -- (5.5,-3pt) node[below]{$V_5$};
\draw (6.3,3pt) -- (6.3,-3pt) node[below]{$V_6$};
\draw (6.7,3pt) -- (6.7,-3pt) node[below]{$V_7$};

\draw[-{Rays[scale=2]} ] (7.2,3pt) node[below, yshift=-.5em]{$U$};

\draw[-{Triangle[scale=2]} ] (3.3,.15) node[below, yshift=-.6em]{$W$};

%------------------- Subject 2 -----------------------%
\draw (-1.5,-1) node{Subject 2};
\draw (0,-1) -- (6.2,-1);
\draw (0,-.9) -- (0,-1.1) node[below,xshift = 10pt,yshift=-5pt]{};
\draw (1.3,-.9) -- (1.3,-1.1) node[below, yshift=-2pt]{$V_1$};
\draw (2.3,-.9) -- (2.3,-1.1) node[below, yshift=-2pt]{$V_2$};
\draw (2.7,-.9) -- (2.7,-1.1) node[below, yshift=-2pt]{$V_3$};
\draw (3.7,-.9) -- (3.7,-1.1) node[below, yshift=-2pt]{$V_4$};

\draw[-{Circle[open, scale=2]} ] (6.3,-.9) node[below, yshift=-.6em]{$C$};

\draw[-{Triangle[scale=2]} ] (4.6,-.9) node[below, yshift=-.6em]{$W$};

%------------------- Subject 3 -----------------------%
\draw (-1.5,-2) node{Subject 3};
\draw (0,-2) -- (10,-2);
\draw (0,-1.9) -- (0,-2.1) node[below,xshift = 10pt,yshift=-5pt]{};
\draw (1.6,-1.9) -- (1.6,-2.1) node[below, yshift=-1pt]{$V_1$};
\draw (6.7,-1.9) -- (6.7,-2.1) node[below, yshift=-1pt]{$V_2$};
\draw[-{Triangle[scale=2]} ] (0,-1.9) node[below, yshift=-.6em]{$W$};

%------------------- Subject 4 -----------------------%
\draw (-1.5,-3) node{Subject 4};
\draw (0,-3) -- (6.5,-3);
\draw (0,-2.9) -- (0,-3.1) node[below,xshift = 10pt,yshift=-5pt]{};
\draw (1.2,-2.9) -- (1.2,-3.1) node[below, yshift=-1pt]{$V_1$};
\draw[-{Rays[scale=2]} ] (6.5,-2.9) node[below, yshift=-5pt]{$U$};

%------------------- Subject 5 -----------------------%
\draw (-1.5,-4) node{Subject 5};
\draw (0,-4) -- (3.2,-4);
\draw (0,-3.9) -- (0,-4.1) node[below,xshift = 10pt,yshift=-5pt]{};
\draw[-{Circle[open, scale=2]} ] (3.3,-3.9) node[below, yshift=-.6em]{$C$};

\end{tikzpicture}}  
     \caption{Example patient trajectories. Triangle depicts time of treatment switching $W$, a pipe $|$ indicates time of $k^{th}$ occurrence of recurrent event at time $V_k$, an $\times$ indicates death at time $U$, and $\circ$ indicates censoring event at time $C$. Subjects 1 and 2 initiate treatment within followup, Subject 3 initiates at the beginning of followup, and Subjects 4 and 5 never initiate. Subject 3 reaches the end of the followup window, leading to complete information on survival and event count. Similarly, information is complete for subjects who die within followup since events cannot occur after death. Survival and recurrent event information is missing for Subjects 2 and 5 who are censored before $\tau$. \bl{The variables $\tilde C = \min(C, \tau)$, $\tilde U = \min (U, \tilde C)$, and $\tilde W = \min (W, \tilde U)$ are transformations of the variables in this figure representing a subject's censoring time, end of followup (due to first of either death or censoring), and first of either treatment switching time or end of follow-up, respectively. The dashed lines show the partition of the followup $(0,\tau]$ into $k=1,2,\dots, K=10$ intervals given by $I_k = (\tau_{k-1},\tau_k]$ as described in Section \ref{sc:obs_data}. Subject 1, for example, was not alive in interval 9 and 10 and thus has survival trajectory $\bar T_{10} = (0,0,0,0,0,0,0,0,1,1)$. Since they were uncensored, $\bar C_{10} $ is the length-10 zero vector. Since they switched to treatment in interval 4, $\bar A_{10} = (0,0,0,1,1,1,1,1,1,1)$. Finally, their recurrent event trajectory is given by $\bar Y_{10} = (0,1,1,1,1,1,2,0,0,0)$. }
 }
     \label{fig:diag}
\end{figure*}

\section{Observed Data Structure and Discretization} \label{sc:obs_data}
We motivate the setup within the context of our data application for concreteness. The treatment of interest is opioid use and the target population consists of treatment-naive patients newly diagnosed with chronic pain. Eligibility is defined using two diagnosis codes for chronic pain and no claims for opioids in the year prior to diagnosis. Upon the first instance of meeting all the eligibility criteria, the patient is considered to be eligible for treatment. A schematic of such data is given in Figure \ref{fig:diag}. At the time of eligibility (time zero), we record a set of covariate information, denoted by $p-$dimensional vector $L \in \mathcal{L}$, \bl{and follow patients for a maximum of time $\tau$ (measured in weeks)}. Though any covariate type is allowable, we assume all covariates are discrete to avoid measure-theoretic notation in the presentation. In the observed claims data, patients with chronic pain initiate opioid within the followup period at time $0\leq W < \tau$. For patients who switch to opioids exactly at eligibility, $W=0$, while for patients who are never assigned opioids in the followup period $W>\tau$. Following eligibility, patients may die at time $U$ or be censored at time $\tilde C = \min(C,\tau)$, where $C$ is some subject-level censoring time. In our Medicare claims data, $C$ is the first of either the date of the end of the data cut or date at which medicare eligibility is lost - both relative to time zero. The value $\tau>0$ is the end of the followup window of interest measured from time zero. Let $\tilde U = \min(U, \tilde C)$ be the observed time on study with death indicator $\delta_U = I(U<\tilde C)$. Additionally, we observe event occurrence history up to this time. Specifically, we observe $J$ events with the $j^{th}$ event occurring at time $V_{j} < \tilde U$. In practice, patients may die or be censored before we ever observe treatment switch. \bl{Let $\tilde W = \min (W, \tilde U)$ }be the potentially censored time of treatment switch with indicator $A = I(W < \tilde U)$ being one if a switch was observed during followup and zero otherwise. \bl{We observe these variables for $n$ independent subjects indexed by $i=1,2,3,\dots, n$ so that the observed data is $D = \Big \{\tilde W_i, A_{i}, L_i, \tilde U_i, \delta_{Ui}, \{ V_{ij} \}_{j=1}^{J_i}, \Big \}_{i=1}^n$. When referring to variables for an arbitrary subject $i$ we avoid indexing for compactness.}

Let $P(D)$ denote the joint distribution of the observed data. Though observed continuously, we discretize the data by partitioning the followup window $[0,\tau)$ into $K>1$ equal-length intervals $I_k = [\tau_{k-1}, \tau_k)$ for $k=1,2, \dots, K$ where $\tau_0=0$ and $\tau_K=\tau$. We define survival status at the beginning of interval $k$ as $T_k = I(\tilde U \leq \tau_{k-1}, \delta_U = 1)$. This indicator process is monotone - taking values $T_k=0$ up to the interval in which death occurs, and then takes values $T_k=1$ afterward. Similarly, let $C_k = I(\tilde U \leq \tau_{k-1}, \delta_U=0)$ be censoring status at the beginning of interval $k$, which is also monotone zero until the interval in which censoring occurs, and then one afterward. We define a time-varying indicator $A_k = I(\tilde W \leq \tau_k, A=1)$ of whether a treatment switch occurred by the end of interval $k$ - the indicator is zero for all intervals before treatment and is one for all $k$ on and after treatment. Finally we define the number of hospitalization events in interval $k$ as $Y_k = \sum_{j=1}^J I(V_{j} \in I_k )$. By convention $T_1 = C_1 = 0$ - that is, patients enter the first interval alive and uncensored.

\section{Potential Outcomes and Average Potential Incidence Rate}
In this section we define the causal incidence rate estimand in terms of potential outcomes. Throughout, we use overbar notation to denote process history, e.g. $\bar X_k = (X_1, X_2, \dots, X_k)$ and underbar notation to denote process futures, e.g. $\underline X_k = (X_k, X_{k+1}, \dots X_K)$. Lowercase letters denote realizations while uppercase letter denote random variables. We define a treatment sequence $\bar A_K = (A_1, A_2, \dots, A_K)$. Recall that the treatment process $\bar A_K$ is monotone - being zero for all $k$ before the interval of treatment and taking on values of 1 for all subsequent points. The vector $\bar A_K$ has possible realizations $a(s) = (\bar a_{s-1}=0, \underline{a}_{s}=1) $, denoting the corresponding length $K$ treatment sequence that is zero up through the interval in which the switch occurs, $s$, and one thereafter. Those who were assigned treatment immediately at eligibility ($s = 1$) followed observed sequence $a(1) = (1,1,\dots, 1)$. Those who never switched ($s=K+1)$ followed sequence $a(K+1) = (0,0,\dots, 0)$. 

We let $T_k^{a(s)}$ denote potential death indicator had a subject, possibly counter to the fact, been following sequence $a(s)$ up to $k$. It is a slight abuse of notation since the index does not contain all components of $a(s)$, but just the components of $a(s)$ up to $k$ - e.g. $T_2^{a(2)} = T_2^{(0,1)}$ while $T_3^{a(2)} = T_3^{(0,1,1)}$. $T_1:=0$ is not a potential outcome since it is definitionally zero for patients who entered the study. Similarly, we define potential number of events in interval $k$, $Y_k^{a(s)}$. Since death is terminal, $P(T_j^{a(s)} = 1 \mid  T_{j-1}^{a(s)} = 1)=1$ and $P(Y_j^{a(s)} = 0 \mid T_{j}^{a(s)} = 1) = 1$ for all $j$. The random vector $\bar T_K^{a(s)}$ takes values $t(k)$ which, like $a(s)$, lives in the space  of binary, monotone, length $K$ vectors. There are $K$ such vectors, which we denote by $t(k) = (\bar t_{k-1}=0, \underline t_{k}=1)$ for $k=1,2,\dots, K$. Similarly, for a subject who dies in interval $k-1$, $\bar Y_k^{a(s)}$ takes values $y(k) = (\bar y_{k-1}, \underline 0_k)$ - length $K$ vectors with non-negative integer values up to and include entry $k-1$ and zero afterwards. A table summarizing this notation is given at the beginning of the supplement.

We target marginal estimands that contrast quantities of the form $E[g( \bar Y_K^{a(s)}, \bar T_K^{a(s)} )]$, where $g$ is some function of the potential outcomes. Our motivating example is a contrast of potential incidence rate, $g( \bar Y_K^{a(s)}, \bar T_K^{a(s)} ) = \frac{ \sum_{k=1}^K Y_{k}^{a(s)} }{K - \sum_{k=1}^K T_{k}^{a(s)} }$, had everyone in the target populations switched at time $s$ versus switched at time $s'$, $\Psi(s, s') = E\Big[ \frac{ \sum_{k=1}^K Y_{k}^{a(s)} }{K - \sum_{k=1}^K T_{k}^{a(s)} } \Big] - E\Big[ \frac{ \sum_{k=1}^K Y_{k}^{a(s')} }{K - \sum_{k=1}^K T_{k}^{a(s')} } \Big]$
%\begin{equation} \label{eq:target}
%    \Psi(s, s') = E\Big[ \frac{ \sum_{k=1}^K Y_{k}^{a(s)} }{K - \sum_{k=1}^K T_{k}^{a(s)} } \Big] - E\Big[ \frac{ \sum_{k=1}^K Y_{k}^{a(s')} }{K - \sum_{k=1}^K T_{k}^{a(s')} } \Big]
%\end{equation}
In each term, the numerator is the potential number of hospitalization events in the followup. The denominator is the potential number of intervals at risk for hospitalization. Some special cases include: 1) $\Psi(1, K+1)$ is the difference in event rates had everyone switched to treatment immediately versus never switched. 2) $\Psi(s, 1)$ is the estimated incidence rate had everyone switched at time $s$ versus switched immediately. 3) $\Psi(s, K+1)$ is the difference in rates had everyone switched at $s$ versus never switched. The estimand captures the fact that potential event rates are composite estimands of person-time (denominator) and event counts (numerator) - and so both processes must be considered jointly as potential outcomes of treatment. Information about survival is not lost since two hypothetical strategies with the same total potential event count - but different potential survival profiles - will have different event rates. Alternatively, comparisons only on the basis of events, $\{Y_k^{a(s)}\}_{k=1}^K$, would lose information on survival: the hypothetical strategies would be found to be equivalent, even though one leads to higher survival. Since our approach models the full transition process, other functions $g()$ can be accommodated. Moreover, indexing the estimand by $s$ can help us assess the effect of timing. As is standard in causal inference, we handle censoring as an additional intervention along with treatment \citep{Robins2001}. Potential outcomes are implicitly indexed by joint intervention of setting switching time to $s$ \textit{and} eliminating censoring,  $(\bar c_K =0, a(s))$. We avoid explicit indexing to reduce notational burden.

\subsection{Causal Identification under Right-Censoring}
For some $s$, $E[g( \bar Y_K^{a(s)}, \bar T_K^{a(s)} )]$ is an average over the joint probability mass function (pmf) of the potential outcomes, $P^*(\bar Y_K^{a(s)} = y(k), \bar T_K^{a(s)}= t(k))$ - this includes each term of $\Psi(s, s')$. Since this is a distribution of unobserved counterfactuals, to identify $P^*$ in terms of the observed-data distribution we require that several causal identification assumptions hold:
\begin{enumerate}
    \item[1.] \textbf{Sequential Ignorability}:
    $$ \underline Y_{k}^{a(s)}, \underline T_{k}^{a(s)} \perp C_k, A_k \mid \bar A_{k-1}, L, \bar Y_{k-1},  C_{k-1} = T_{k-1} = 0$$
\end{enumerate}
    for each $k$. That is, among those at risk for an event at interval $k$, treatment and censoring at that interval are unrelated to future potential outcomes conditional on available history. This rules out situations in which patients who switch at interval $k$ are those who would have systematically had higher outcome incidence relative to those who did not, even after adjusting for history $(\bar A_{k-1}, L, \bar Y_{k-1})$. It also rules out situations in which patients censored at interval $k$ would have gone on to have different event incidence relative to those who are uncensored even after adjusting for available history. To identify outcomes under a hypothetical switch at interval $s$, it must be possible to remain uncensored, alive, and untreated until interval $s-1$ and then switch at interval $s$. So, we require
\begin{enumerate}
    \item[2.] \textbf{Switching Positivity}: For each $l$ with $f_L(l)>0$,
    \begin{equation} \label{eq:positivity}
    \begin{split}
    \lambda_s^A(l, & \bar y_{s-1} ) ( 1 - \lambda_s^C(l, \bar y_{s-1}) ) \times  \\ 
    & \prod_{k=1}^{s-1}  (1-\lambda_k^A( l, \bar y_{k-1} ))( 1 - \lambda_k^C(l, \bar y_{k-1} )) > 0
    \end{split}
    \end{equation}
\end{enumerate}
Above, $ \lambda_k^A( l, \bar y_{k-1} ) = P(A_k = 1 \mid \bar A_{k-1} = T_{k-1} = C_{k-1}=0, l, \bar y_{k-1})$ is the discrete-time cause-specific hazard of treatment and $\lambda_k^C(l, \bar y_k) =  P(C_k = 1 \mid \bar A_{k-1} = T_{k-1} =C_{k-1}=0, l, \bar y_{k-1}) $ is the discrete-time cause-specific hazard of censoring. Additionally, it must be possible to remain uncensored through the end of follow-up. That is, for each $k=s+1, s+2, \dots, K $, $ ( 1-\lambda_k^C(l, \bar y_k)) >0 $. Violations of either condition precludes estimation of incidence rates over the desired followup window without parametric smoothing. A stable unit treatment value assumption (SUTVA) is also required at each interval to connect observed and potential outcomes: at each interval $k$, $Y_k = \sum_{s} Y_k^{a(s)} I( \bar A_k = a(s), \bar C_k = 0_k )$ and  $T_k = \sum_{s} T_k^{a(s)} I( \bar A_k = a(s), \bar C_k = 0_k )$. As before, due to difference in dimension, $ \bar A_k = a(s)$ is a minor abuse of notation as it denotes equality only up to $k^{th}$ entry.

Under these assumptions we can identify $P^*$ via the g-formula
\begin{equation} \label{eq:gform} 
    \begin{split}
        P^*\Big(y(k), & t(k)\Big) 
         = \sum_{l\in \mathcal{L}} \lambda_k(a_k^s, \bar y_{k-1},l)  \times \\
        & \prod_{j=1}^{k-1} f(y_j \mid a_j^s, \bar y_{j-1}, l )\big(1 - \lambda_j(a_j^s, \bar y_{j-1},l)\big)f_L(l)
    \end{split}
\end{equation}
with support in $k=1, 2,\dots, K+1$ and switching indicator being set to $a_k^s=I(k>s)$. That is treatment is zero up until interval $s$ and then one afterward. A proof is provided in Supplement S1. There are three unknowns in this expression which can be learned from observed data: 1) A model for the discrete-time death hazard, $ \lambda_k(a_k, \bar y_{k-1}, l) = P(T_k = 1 \mid T_{k-1}= C_{k-1}=0 , a_k, \bar y_{k-1}, l ) $ conditional on treatment history, event history, and covariates, which governs the terminal event process; 2) A model for the distribution of the number of events in interval $k$, given survival up to that interval $ f(y_k \mid a_k,\bar y_{k-1}, l ) = P(Y_k=y_k \mid T_k = C_k=0, a_k, \bar y_{k-1}, l) $ conditional on treatment, event history, and covariates - which governs the recurrent event process; 3) A model for the joint distribution of the covariates $f_L(l)$. Bayesian inference follows from obtaining a posterior distribution over the parameters governing these models - which induces a posterior on the joint, $P^*$, and functionals of it such as $\Psi(s, s')$. The required averaging will be done by simulation in a g-computation procedure.

\subsection{Timing Misalignment, Grace Periods, and Positivity Violations}

Here we provide a formal causal framing of various naive methods, the conditions under which they are used, and the issues that arise. Issues related to timing misalignment arise when we would like to estimate the effect of some treatment within a target population, but relatively few subjects are actually treated exactly at eligibility. From a causal perspective, this is a near violation of sequential positivity, $P(A_1 = 1 \mid l, y_{0}) \approx 0$.  In order to circumvent the issue, it is common to look past eligibility to see if any switch to treatment occurred during the followup. If a patient ever switches to treatment in the followup, i.e. $\max \underline A_1 = 1$, we consider the patient ``treated''. If they do not, $\max \underline A_1 = 0$, then we flag them as never treated and proceed with an  ``ever-versus-never'' comparison. This approach targets $ E[	\frac{  \sum_{k=1}^K Y_{k} }{K - \sum_{k=1}^K T_{k}}  \mid \max \underline A_1 = 1 ]	-	 E[ \frac{  \sum_{k=1}^K Y_{k} }{K - \sum_{k=1}^K T_{k}}  	\mid \max \underline A_1 = 0 ]	$. This is an associational contrast as it does not involve any potential outcomes - it is simply a comparison of incidence rate between those who happened to switch to treatment during the followup and those who did not. It is unclear what (if any) causal contrast this corresponds to and under what assumptions. However, it is clear that this contrast is \textit{conditional } on post-baseline information and so is in general not equal to the \textit{marginal} contrast we define. Conditioning on future information in this way is problematic because it attributes survival and events occurring before the switch to the treatment. Another common alternative is to allow for a certain grace period after eligibility. If a patient switches to treatment within, say, a grace period of 10 weeks, then the patient is flagged as treated. Otherwise, they are flagged as untreated. The approach then proceeds to analyze outcomes after 10 weeks. Explicitly, the associational quantity this targets is given by    $E[	\frac{  \sum_{k=10}^K Y_{k} }{ (K-10) - \sum_{k=10}^K T_{k}} \mid \max \bar A_{10} = 1, \bar{T}_{10} = 0 ]	-	E[ \frac{  \sum_{k=10}^K Y_{k} }{ (K-10) - \sum_{k=10}^K T_{k}}  \mid  \max \bar A_{10} = 0, \bar{T}_{10} = 0  ] $. This is again circumventing a positivity violation: the hope is that, even though too few subjects switched to treatment at baseline, more patients will have switched to treatment by 10 weeks. As before, it is unclear what causal quantity this targets. However, this moves the estimand away from the marginal one we want and towards one that is conditional on survival through an (arbitrary) grace period. In contrast to the above approaches, our approach targets well-defined causal estimands under clearly stated identification assumptions and correctly attributes recurrent events and survival to each treatment status in a time-varying fashion. The estimand $\Psi(s, K+1)$, for example, has a direct interpretation as the marginal difference in average incidence rates had everyone in the target population switched at week $s$ versus never switched. Computing the estimand for several values of $s$ helps gauge the effect of treatment initiation delay of $s$ weeks. Our causal framing in the previous section clarifies that, when choosing $s$, avoiding sensitivity to model misspecification requires choosing values supported by the data as these are the values for which positivity holds.

\section{Bayesian Semiparametric Models with Shrinkage Priors}
In this section we outline Bayesian models for the terminal event probability, $\lambda_k(a_k, \bar y_{k-1}, l)$, as well as the distribution of recurrent events $f(y_k \mid a_k,\bar y_{k-1}, l )$ at each $k$. Across $k$, these two models characterize the joint evolution of the recurrent event process and the terminating process as a function of treatment and covariates. We model the processes via the following two mean functions
\begin{equation}
    \begin{split} \label{eq:mods}
    \lambda_k(a_k, \bar y_{k-1}, l; \beta ) & = \expit\big(\beta_{0k} + l' \beta_{L} + y_{k-1} \beta_{Y} + \beta_{A} a_k \big) \\
    \mu_{k}(a_k, \bar  y_{k-1}, l; \theta ) & = \exp\big(\theta_{0k} + l' \theta_{L} + y_{k-1}  \theta_{Y} + \theta_{A} a_k \big)        
    \end{split}
\end{equation}
Here, $\lambda_k$ governs the death process. The function $\mu_k$ is the conditional mean of $f(y_k \mid a_k,\bar y_{k-1}, l )$, which can be any distribution with appropriate support for count outcomes such as Poisson, Negative Binomial, or zero-inflated versions of the two. \bl{The time-specific intercepts allows flexibility: as $K$ increases, the partition gets finer and the baseline hazard in the model $\lambda_k$, controlled by intercepts $\{ \beta_{0k}\}_{k=1}^K$, gets more flexible and the model limits to Cox's semiparametric proportional hazard model \citep{Thompson1977}. Similarly, $\mu_{k}$ is a proportional mean model for the recurrent events at interval $k$, with a flexible baseline intensity/event rate controlled by intercepts $\{\theta_{0k}\}_{k=1}^K$. Above, the coefficients $\beta_{Y}$ and $\theta_{Y}$ capture temporal dependence across repeated hospitalization measures per subject over intervals. As is frequently done in causal models, a Markov assumption is invoked above so that dependence on hospitalization history, $\bar y_{k-1}$, is modeled only through the first-order lag, $y_{k-1}$.} In \eqref{eq:mods} we present a linear and additive specification for simplicity but note that interactions, polynomial functions, and basis splines can all be used to incorporate non-linear and non-additive effects. Finally, while we index the models for $k=1,2, \dots, K$ note that for $k=1$ there is no death model since $T_1=1$ for everyone by convention. Additionally, the model for $Y_1$ does not condition on history since there is no history at the first interval. For compactness, we will denote the hazard and event rate parameters as $\beta = ( \{\beta_{0k}\}_{k=1}^K, \beta_L, \beta_Y, \beta_A  )$ and $\theta = ( \{\theta_{0k} \}_{k=1}^K, \theta_L, \theta_Y, \theta_A )$, respectively.

\begin{figure*}
    \centering
    \includegraphics[scale=.45]{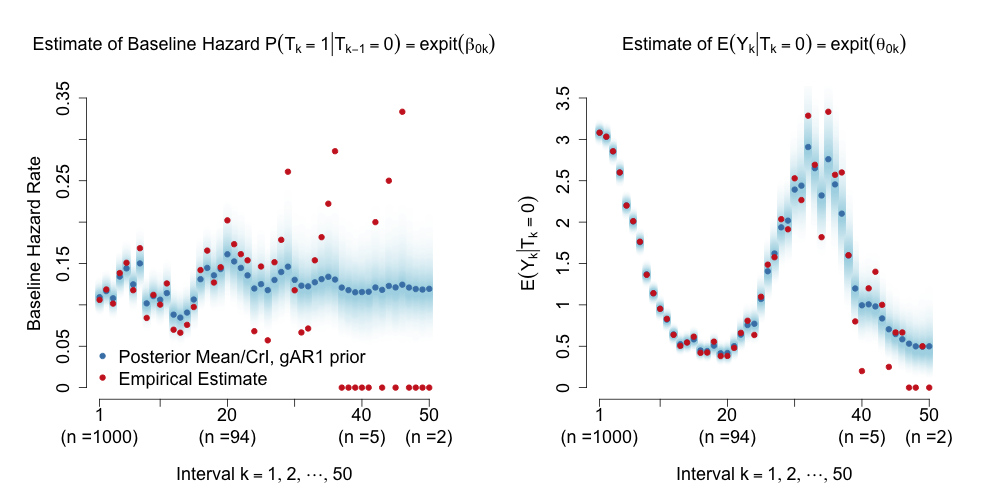}
    \caption{Posterior $gAR1$ smoothing of baseline hazard of logistic regression (left) and baseline mean of Poisson regression (right) demonstrated in a simulated example with at-risk counts on the x-axis label. Blue points and shading depict posterior mean and 95\% credible interval estimates, respectively. At earlier time points with many subjects at risk empirical and Bayesian estimates agree. By interval 35, there are fewer than 10 subjects at risk, leading to erratic empirical hazard  and mean event count estimates. For exmaple, in many intervals the empirical hazard is exactly zero. This is not due to the true hazard actually being zero in this interval, but because there are too few at-risk subjects for reliable estimation. In contrast, the Bayesian estimates remain stable in these intervals due to prior shrinkage.}
    \label{fig:ar1}
\end{figure*}

A key advantage of Bayesian methods here is that frequentist estimation of \eqref{eq:mods} is challenging in intervals with few patients at risk. Estimation of baseline hazards and event rates via time-specific $\beta_{0j}$ and $\theta_{0j}$, respectively, may be infeasible. This often forces ad-hoc simplifications of the model - perhaps assuming a constant intercept $\beta_{0j} = \beta_0$ for each $j$ - in practice. In contrast, we use a smoothing prior process on the time-varying coefficients, $\{\beta_{0k}\}_{k=1}^K$ and $\{\theta_{0k}\}_{k=1}^K$. We say a sequence of random variables, $\{X_k\}_{k=1}^K$, follows a Gaussian first-order autoregressive process if $X_1  = \eta + \sigma \epsilon_1 $ and for $k=2, \dots, K$, $X_k = \eta(1-\rho) + \rho X_{k-1} + \sigma \epsilon_k$. Here, $\epsilon_k \stackrel{iid}{\sim} N(0,1)$ and $\eta$, $\rho$, and $\sigma$ are the parameters of the process. We denote this as $\{X_k\}_{k=1}^K \sim gAR1(\eta, \rho, \sigma)$. The mean of the process is given by $E[X_k] = \eta $ while $Corr(X_{v}, X_{v-u}) = \rho^{u}$, where $-1<\rho<1$, is the correlation between $X_v$ and the state of the process $u$ periods in the past, with correlation declining as we go further back in time. To enforce smoothness in the time-varying coefficients, we specify priors $\{\beta_{0k}\}_{k=1}^K \sim gAR1(\beta_0, \rho_\beta, \sigma_\beta)$ and $\{\theta_{0k}\}_{k=1}^K \sim gAR1(\theta_0, \rho_\theta, \sigma_\theta)$, which induces dependence between the baseline death/event rates across intervals. This is a sensible prior to shrink towards as it reflects the prior belief that - in the absence of data in interval $k$ - the baseline hazard rate shouldn't be too different from the hazard rate in the previous interval, providing useful regularization at later intervals with fewer patients at risk. Figure \ref{fig:ar1} illustrates the smoothing effect of the prior in a synthetic data set. We place weakly informative priors over the hyperparameters, $\rho_\beta, \rho_\theta \sim U(-1,1)$, $\beta_0,\theta_0 \sim N(0, 1)$, with flat $f(\sigma_\beta) \propto 1$ and $f(\sigma_\theta) \propto 1$. For the covariate distribution model we use the nonparametric Bayesian bootstrap \citep{Rubin1981} by setting $f_L(l; \pi ) = \sum_{i=1}^n \pi_i \delta_{L_i}(l)$ where $\pi=(\pi_1, \pi_2,\dots, \pi_n)$ lives in the simplex and follows a conjugate Dirichlet prior $\pi \sim Dir(0_n)$ with resulting posterior is $\pi \sim Dir(1_n)$. Here, $0_n$ and $1_n$ denote the length $n$ vector of zeros and ones, respectively.
\section{Posterior Sampling and Bayesian G-computation}
\label{sc:computation}
Bayesian inference for $\Psi(s,s')$ requires the posterior distribution of the parameters governing the unknown models in the g-formula in \eqref{eq:gform},  $\omega = (\theta, \beta, \pi )$. Markov Chain Monte Carlo (MCMC) methods can be used to obtain, say, $M$ draws from the posterior, indexed by $m=1,2,\dots, M$. In Supplement S2, we provide the likelihood construction and joint posterior up to a proportionality constant which can be fed into standard software such as \texttt{Rstan} in \texttt{R} to obtain posterior draws. The idea of g-computation is to simulate the terminal and event count processes jointly under both interventions. For each parameter draw, the difference in the average incidence rates can then be computed, yielding a draw from the posterior of $\Psi(s,s')$. Specifically, suppose we are given a posterior draw of the parameters $\omega^{(m)} = (\theta^{(m)}, \beta^{(m)}, \pi^{(m)} )$, the observed set of covariate vectors $\{l_i\}_{i=1}^n$, and a desired switching intervention, $s$. 
\begin{table*}[t!]
\centering
\caption{Results across 1000 simulation runs: absolute average bias along with relative bias, variance, and mean-squared error (MSE) of posterior mean estimator of $\Psi(6,K+1)$ with $K=12$ time points and $n=500$ subjects at time zero. Interval width and coverage 95\% credible interval (for Bayes) and 95\% confidence interval formed via percentiles of 500 bootstrap iterations (for other methods).}
\label{tab:simulation results}
\begin{tabular}{ll|ccccccc}
\hline
 &  & \% Bias &  Rel. Bias  & Rel. Var &  Rel. MSE & Interval Width & Coverage \\ \hline

\multicolumn{1}{c}{\multirow{4}{*}{Setting 1}} 
                     & Bayes gAR1          & 0.03   & 1       & 1      &  1  & 6.34   & .96  \\
\multicolumn{1}{c}{} & Freq GLM            & 0.01   & 0.19     & 1.17   & 1.11   &  6.41  &  .98 \\
\multicolumn{1}{c}{} & Grace Period        &  0.12  & 3.75    & 2.17   &  2.83   &  8.16  &  .91 \\
\multicolumn{1}{c}{} & Ever-Never          &  4.75 & 142.94  & 42.85  &  1182.09  & 34.62    &  0 \\ \hline

\multirow{4}{*}{Setting 2} 
                     & Bayes gAR1    &   0.01   & 1       &  1      & 1        &  9.25    & .93  \\
                     & Freq GLM            &   0.03   &  4.14   &  1.27   &  1.29    &  10.99   &  .96\\
                     & Grace Period        &   0.42   &  62.01  &  0.54   &  4.11    &  6.38    &  .25\\
                     & Ever-Never          &   3.26   &  481.83 &  12.79  &  228.40  &  29.58   &  0\\ \hline
\end{tabular}
\end{table*}

\begin{enumerate}
    \item For each subject $i$, we first simulate $b=1,2,\dots, B$ trajectories from the joint terminal and recurrent event processes defined by model in Equation \eqref{eq:mods}, denoted $\{  Y_{ki}^{(b),(m)} , T_{ki}^{(b),(m)} \}_{k=1}^K $, while setting $A_k = I(k \geq s)$. Details are provided in Supplement S4.
    \item  For each subject $i$, a Monte Carlo approximation to the conditional (on $L=l_i$) expected incidence rate is given by
    $ E^{(m)} \Big[   \frac{\sum_{k=1}^K Y_k^{a(s)}}{K - \sum_{k=1}^K T_{k}^{a(s)} }  \mid L = l_i \Big] \approx \frac{1}{B}\sum_{b=1}^B \frac{\sum_{k=1}^K Y_{ki}^{(b),(m)}}{K - \sum_{k=1}^K T_{ki}^{(b), (m)} } $
    \item Repeat Steps 1-2 under the second intervention $s'$.
    \item Using a draw of the Bayesian bootstrap weights, $\pi^{(m)} = (\pi_1^{(m)}, \pi_2^{(m)}, \dots, \pi_n^{(m)}) \sim Dir(1_n)$ we average the difference in conditional expectations over the covariate distribution
   \begin{equation*}
   	\begin{split}
	\Psi(s,s')^{(m)} & \approx \sum_{i=1}^n\Big\{  E^{(m)} \Big[   \frac{\sum_{k=1}^K Y_k^{a(s)}}{K - \sum_{k=1}^K T_{k}^{a(s)} }  \mid L = l_i \Big] \\ 
	& \ \ \ -  E^{(m)} \Big[   \frac{\sum_{k=1}^K Y_k^{a(s')}}{K - \sum_{k=1}^K T_{k}^{a(s')} }  \mid L = l_i \Big] \Big\} \pi_i^{(m)}  
	\end{split}
   \end{equation*} 
\end{enumerate}
Doing this for each posterior draw of the parameters, yields a set of $M$ posterior draws from the causal contrast of interest, $ \{ \Psi(s,s')^{(m)} \}_{m=1}^M$. The posterior can be summarized in all of the usual ways. For instance, the posterior mean can be approximated by taking the average of the draws. An equal-tailed $(1-\alpha)100\%$ credible interval can be formed by taking the 2.5th and 97.5th percentiles of these draws. Inference in the frequentist paradigm follows the same procedure as above, but $m=1,2,\dots, M$ will index parameters estimating using the $m^{th}$ bootstrap re-sample of the full data. In large samples, estimators will perform similarly, but the smoothness of the Bayesian approach is desirable in sparse settings when we encounter convergence issues in the frequentist bootstrap. In general Monte Carlo approximation error declines at a rate of $1/\sqrt{B}$ and across simulation settings a value of $B=50$ was sufficient to achieve good performance in simulations. As practical guidance in any given application, for a given posterior draw, one can repeat steps 1-4 for successively larger $B$. Once approximations are similar, at say, the third decimal place (or other level of desired precision) we can stop and use that value of $B$ across draws. \bl{Other Monte Carlo methods tailored for such causal inference problems may also be considered \citep{Linero2022}.}

\section{Simulation Studies Assessing Finite-Sample Performance}

Though Bayesian in formulation, the prior shrinkage of the proposed approach may lead to posterior causal estimates with good frequentist properties especially in sparse conditions. Finite-sample properties of the proposed models have not been previously assessed so it is of interest to verify when frequentist properties are acceptable. Thus, we designed a simulation study in which 1,000 data sets were generated with $n=500$ subjects each. Discrete-time longitudinal survival and recurrent events were simulated for $K=12$ intervals. Survival at each interval was simulated from a logistic model conditional on treatment and a set of five baseline covariates (two binary and three continuous). The number of events in each interval was simulated from a Poisson distribution with conditional mean dependent on treatment, baseline covariates, and number of events in the previous interval. We simulate data in two settings. In Setting 1, censoring is relatively light about 190/500  subjects were censored before the last interval, on average across the 1,000 simulated datasets. About 180 patients either died while the remaining survived uncensored through the end of the followup - leaving us with a decent amount of of subjects with complete survival/event rate data within the followup. In Setting 2, censoring rates were higher with 335/500 patients being censored in a typical data set. There were only about 85/500 deaths and only about 80 subjects who survived through the end of the followup uncensored within a typical simulated dataset. Relative to Setting 1, much fewer subjects have complete event/survival information throughout the followup. In both setting, censoring status in each interval was simulated conditional on baseline covariates. More details on the simulation study setup are provided in Supplement S3.

\begin{table*}[t!]
\centering
\caption{Across columns, $N$ is the count of patients at risk by a given week. Rows summarize baseline characteristics among those at risk. Notably, average Gagne score declines over time as patients with higher baseline Gagne scores either drop out or die. This motivates adjustment for Gagne score.}
\label{tab:table1}
\begin{tabular}{l|llll}
 & Week 1 & Week 13 & Week 26 & Week 39 \\ \hline
 & N =1391 & N = 1341 & N = 1254 & N = 1139 \\ \hline
Male & 857 (61.6\%) & 827 (61.7\%) & 767 (61.1\%) & 699 (61.3\%) \\
Age & 63.3 (16.2) & 63.3 (16.1) & 63.2 (16.1) & 63.1 (16.1) \\
Region &  &  &  &  \\
\hspace{3mm} Northeast & 365 (26.2\%) & 344 (25.6\%) & 314 (25.0\%) & 283 (24.8\%) \\
\hspace{3mm} Midwest & 402 (28.9\%) & 389 (29.0\%) & 355 (29.2\%) & 338 (29.8\%) \\
\hspace{3mm} West & 275 (19.7\%) & 268 (19.9\%) & 252 (20.0\%) & 225 (19.7\%) \\
\hspace{3mm} South & 349 (25.1\%) & 340 (25.4\%) & 323 (25.8\%) & 293 (25.7\%) \\
Race &  &  &  &  \\
\hspace{3mm} Black & 60 (4.3\%) & 57 (4.3\%) & 53 (4.2\%) & 46 ( 4.0\%) \\
\hspace{3mm} White & 1227 (88.2\%) & 1182 (88.1\%) & 1104 (88.0\%) & 1007 (88.4\%) \\
\hspace{3mm} Other & 104 (7.5\%) & 102 (7.6\%) & 97 (7.7\%) & 86 (7.6\%) \\ 
Gagne Score & 1.33 (1.9) & 1.32 (1.9) & 1.27 (1.9) & 1.25 (1.8) \\ \hline
\end{tabular}
\end{table*}

For each simulated data set we apply several methods to estimate the causal effect of switching by interval 6, $\Psi(6, 13)$. The results of the simulation are presented in Table \ref{tab:simulation results}. We use the Bayesian models with the gAR1 prior described in the previous section (labeled ``Bayes gAR1'') and the frequentist analogue (labeled ``Freq GLM''). For the Bayesian models, we use uninformative priors described in Supplement S3 and in the previous sections. In order to isolate the effect of the shrinkage on performance, both models are correctly specified. In Setting 1, both models have relatively low bias as a percentage of the true effect - 3\% for Bayes gAR1 versus 1\% for Freq GLM. While average finite-sample bias of the posterior mean estimator is slightly higher for the Bayesian model, the variance is lower due to the regularizing gAR1 prior - this leads to an overall lower mean squared error (MSE). This aligns with usual bias-variance trade off that comes with Bayesian and frequentist estimators. In terms of interval estimation, the Bayesian method yields a slightly narrower interval with closer to nominal coverage.

As comparators we also include two common methods often applied in practice. This is mainly to highlight the importance of proper joint modeling of recurrent events and the deficiencies of ad-hoc approaches. The first is the Grace Period method where we assign subjects to treatment based on whether or not they have switched to treatment by interval six. Event rates are then estimated using data after interval six using a Poisson regression with observed time on study as an offset and the total number of events as the outcome. Predictions from these Poisson models are obtained for each subject under both treatment assignments and the difference is averaged to obtain a point estimate of the incidence rate difference. A bootstrap procedure with 500 resamples is used to compute interval estimates. In Setting 1, this approach yields a biased estimate with average relative bias of around 12\%. This is due in part to the exclusion of subjects who died or censored before reaching interval 6. The other naive method is the ``ever-never'' method which simply classifies patients as treated ($A=1$) if they ever switched to treatment in the follow up and untreated ($A=0$) if they never switched. A Poisson regression was used to estimate event rates with the total number of events in the followup period as the outcome and the total observed followup time included as an offset. This approach is biased due to the misattribution of person-time and event rates. In Setting 2, the data are much more sparse due to heavier censoring. This is where the induced shrinkage of the Bayesian gAR1 method provides a significant advantage. While both the frequentist and Bayesian methods have close to nominal coverage, the Bayesian method yields a narrower interval. Moreover, finite-sample bias, variance, and MSE are all lower for the Bayesian method. This highlights the benefit of shrinkage in sparse data settings. The relative performance of the two naive methods worsens as censoring rate increases. In Supplement S3, we provide additional results in settings with more intervals ($K=52$ as in our data application) and model misspecification. \bl{The results are quite similar to those reported here, however with model misspecification the gAR1 credible intervals exhibit below nominal coverage in certain settings.}

\section{Analysis of Opioid use Among Patients with Chronic Pain}

According to 2016 estimates, 20\% of adults in the United States suffer from chronic pain \citep{dahlhamer2018}. While opioids are commonly used to manage pain, approximately 8\%-12\% \citep{vowles2015} of patients with chronic pain also have opioid use disorder (OUD), diagnosed based on unsuccessful attempts at controlling opioid use. The effects of prescribing opioids among chronic pain patients is an active research question. In this work we focus on acute and critical care hospitalization outcomes among patients with chronic back pain who have a history of OUD at the time of chronic pain diagnosis. On the one hand, prescribing opioids may potentially exacerbate OUD and lead to increased hospitalization rate. On the other hand, failing to prescribe opioids may cause patients to subsequently seek illicit opioids which could lead to hospitalizations. Previous work has described the need to understand the relationship between opioids and hospitalization risk as it can be used to ``inform the development of tailored interventions and policy recommendations''  \citep{Moyo2022, Moyo2024}. However, a formal causal treatment of the problem remains lacking.

We analyze data from a nationally representative 20\% random sample of Medicare claims data from January 1, 2016 to December 31, 2019. A key eligibility criterion is a diagnosis of chronic back pain, defined by two claims diagnosis codes for back pain separated at least three months apart. Additionally, we require that patients have been enrolled in Medicare for 12 months before diagnosis and have had no opioid prescriptions over that time. Followup (time zero) begins at the first instance that a patient meets these eligibility criteria. Since we are interested in evaluating effects among patients with OUD, we subset to those who have a history of OUD at time zero. There are a total of $N=1391$ patients with OUD who meet these criteria. At time zero, we record values of baseline characteristics summarized in Table \ref{tab:table1}, which we collectively denote as $L$. Of particular importance is the Gagne Score which is a composite score of comorbidities like cancer and liver disease which may be related to both opioids and hospitalization risk. Higher score values indicate greater comorbidity and mortality risk. We also record whether they are prescribed opioids at the time of diagnosis. We count outcomes from time zero for 1 year (the followup window), death, or censoring - whichever comes first. About 24\% of patients were censored within the 1-year followup. This was mostly administrative due to end of data cut at December 31, 2019 (about 65\% of censored patients). The remaining patients were censored due to to loss of Medicare eligibility. We partition the followup period into $K=52$ weekly intervals index by $k$. In each interval, we record whether a patient is alive or dead entering that interval ($T_k$), if $T_k=0$ then we record the number of hospitalizations in that interval, $Y_k$. We also record whether a patient is censored upon entering interval $k$, $C_k$. Additionally we record whether a patient has received an opioid treatment by week $k$, $A_k$. Only 25 patients ($\approx 2\%$ of the sample) switched to opioid exactly at time zero while 303 subjects ($\approx 22\%$) switched within one year. The remaining 1063 subjects ($\approx 76\%$ of the sample) never switched within followup. 

\begin{figure*}[t]
    \centering
    \includegraphics[scale=.3]{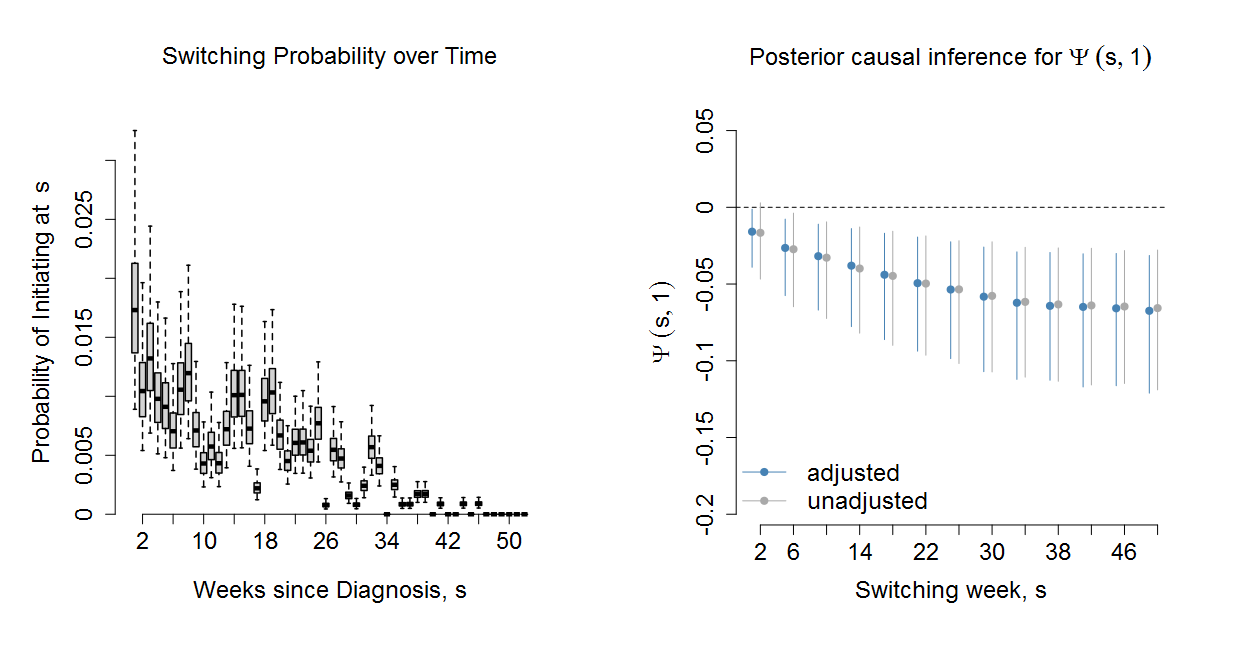}
    \caption{Left: Checking positivity. Each boxplot shows the distribution of estimated probability of switching to opioid at week $s$. Distributions centered above zero at a given $s$ indicates positivity holds for that $s$. For example, very few patients switch to opioids after week 34 and boxplots tend to be right at the bound. So, estimation of $\Psi(s,1)$ is sensitive to extrapolation error for $s\geq 34$. Right: Posterior mean and 95\% credible interval estimates of the causal incidence rate difference $\Psi(s, 1)$ - the difference in event rates had everyone switched to opioids $s$ weeks after diagnosis versus immediately at diagnosis. The units are hospitalizations per year. Delaying opioids tends to reduce hospitalization rates. For example, starting opioids 26 weeks after chronic pain diagnosis leads to about .05 fewer hospitalizations per year.}
    \label{fig:positivity}
\end{figure*}

The target estimand is the average difference in hospitalization rates had everyone switched to opioids $s$ weeks after diagnosis versus immediately switched to opioids denoted by $\Psi(s,1)$. Negative values indicate lower hospitalization risk due to a delay of $s$ weeks while positive values indicate higher hospitalization risk. We assume the required identification conditions such as sequential ignorability and positivity hold. Sequential ignorability is fundamentally untestable and we cannot rule out violations due to unmeasured confounding. However, we do condition on a set of measured baseline covariates such as age, sex, race, and gagne score - which are known drivers of both opioid prescribing and OUD. Additionally, we allow opioid treatment at each time point to depend on hospitalization history up to that time point as well as time since chronic pain diagnosis. Thus, sequential ignorability is weaker than naive approaches that assume ignorability conditional on only time-fixed baseline covariates. We feel somewhat confident that censoring is also ignorable since it is largely administrative. Unlike ignorability, positivity is testable as we can directly estimate the cause-specific hazards of treatment and censoring and check for which $s$ they are bounded. This will help determine for which $s$ the estimation of $\Psi(s,1)$ relies on model extrapolation. To assess positivity, we model the discrete-time hazards via logistic regressions and use maximum likelihoood estimation to obtain estimated hazards $\hat \lambda_k^A(l_i, \bar y_{k-1,i})$ and $\hat \lambda_k^C(l_i, \bar y_{k-1,i})$ for each subject. Using these models, we can then compute Equation \eqref{eq:positivity} for each subject $i$, and then plot the distribution of these probabilities for the various $s$ of interest. More details are provided in supplement S6. The left panel of Figure \ref{fig:positivity} depicts the distribution for each week using boxplots. Notice that the probability of switching to opioid at week 34 is nearly zero. This means that estimates of $\Psi(34,1)$ rely on model extrapolation and therefore sensitive to model misspecification. Relatively few patients switched to treatment exactly at any particular week, implying that parametric smoothing is unavoidable. Similarly, recall that in order to estimate $\Psi(s,1)$ there must be some patients who survive uncensored through the end of the follow up. Accordingly, we verified that $(1- \hat \lambda_k^C(l, \bar y_{k-1})) \approx 1$ for all $k$ across subjects.

With this discussion of assumptions in mind, we specify the following model for death $\lambda_k(a_k, \bar y_{k-1}, l; \beta )  = \expit\big(\beta_{0k} + \beta_{A}a_k + \beta_{L}'l + \beta_Y I(y_{k-1}>0) \big)$ and the following for hospitalization risk $\mu_{k}(a_k, y_{k-1}, l; \theta )  = \exp\big(\theta_{0k} + \theta_{A}a_k + \theta_{L}'l + \theta_Y I(y_{k-1}>0) \big)$. In the Supplement S7, we present results from an alternative model that allows the conditional effects $\theta_{A}$ and $\beta_A$ to vary over time. These time-varying coefficients, $\beta_{Ak}$ and $\theta_{Ak}$, are given separate $gAR1$ priors as well to enforce smoothness. The results, while nearly identical to the results in the main text, have wider credible intervals which reflect the increase in number of parameters being estimated.

We use Hamiltonian Monte Carlo Markov Chain (MCMC) as implemented in Stan \citep{carpenter2017} to obtain $M=1000$ posterior draws of all model parameters after discarding 1000 burn-in draws. Then, for each, $s=2,6,10,14, \dots, 50$ (13 values in total), we perform g-computation as described in Section \ref{sc:computation} to obtain estimates of $\Psi(s, 1)$ displayed in the right panel of Figure \ref{fig:positivity}. The scale is number of hospitalizations per year and we see that, generally, opioid initiation delay leads to a reduction in hospitalization risk. For instance, switching to opioids 26 weeks after diagnosis versus immediately leads to about .05 hospitalizations per year with a posterior 95\% interval that excludes zero. This is a relatively small difference and, in fact, the absolute risk of hospitalization is low.

\section{Discussion}

We developed a causally principled, all-in-one method for analyzing recurrent event outcomes with observational data that accounts for several unique complexities at once: treatment misalignment, right-censoring, and termination due to death. From a causal perspective we showed that both time at risk and event counts are downstream of treatments and so must be treated as potential outcomes of treatment itself. We formulate a tailored causal estimand that is broadly applicable in recurrent event analysis and show that, in terms of identification, recurrent event history essentially acts as a time-varying confounder of treatment switching. We construct Bayesian models with smoothing priors and demonstrate in simulations that the resulting causal effect estimators have desirable frequentist properties. Aside from hospitalization history, the method focuses on adjusting for time-constant covariates. Adjustment for time-varying covariates can be easily incorporated but requires modeling the evolution of the these covariates over time along with death and event counts. While we assume sequential ignorability holds, this is fundamentally untestable and it can be violated due to unmeasured confounding.  Although we use g-computation to appropriately adjust for \textit{measured} confounding, without randomization we cannot rule out unmeasured confounding. Future work developing Bayesian sensitivity analyses methods for unmeasured confounding in this setting would be useful. Moreover, we note that the discrete-time approach used here is quite popular in causal inference since the models can be fit using standard software. In these models, we view discretization choice as essentially an implicit smoothing prior with a corresponding bias-variance tradeoff. The Bayesian approach here is a principled way of addressing this: we choose a fine partition but with a prior that smoothes over it. Future work, however, should investigate the utility of continuous-time models that avoid such discretization altogether.

\backmatter

\section*{Acknowledgements}
This work is partially funded by National Institute on Drug Abuse grant R03DA051778 and Patient-Centered Outcomes Research Institute (PCORI) contract ME-2023C1-31348.\vspace*{-8pt}

\section*{Supplementary Materials}
Web Appendices referenced in the manuscript and implementation code are available at the Biometrics website on Wiley Online Library. \vspace*{-8pt}

\section*{Data Availability}
This paper used Medicare Research Identifiable Files accessed through a data use agreement (DUA) with the Centers for Medicare and Medicaid Services (CMS). Our DUA with CMS does not allow sharing of the study data. Researchers interested in using Medicare data can make a request with the Research Data Access Center at www.resdac.org. 

\bibliographystyle{biom} 
\bibliography{mybibilo.bib}

\label{lastpage}

\end{document}